\documentstyle{article}

\catcode`@=11
\newif\ifIncludeFigs\IncludeFigsfalse

\IncludeFigstrue

\ifIncludeFigs
      \@input{epsf} %
      \ifx\epsfbox\undefined
         \EPSFfalse
	 \IncludeFigsfalse
      \else
      \fi
\else
\fi

\ifx\endfloat\notincluded\else
  \nomarkersintext
\fi

\ifx\@ove@rcfont\undefined
  \ifx\citen\undefined
     \def\citen#1{\begingroup \def\@cite##1##2{{##1}}%
	\@citex[]{#1}\endgroup}
  \fi
\else
  \def\@cite#1{$\@ove@rcfont\m@th^{[{#1}]}$}
\fi

\def\caption{\@dblarg\aux@caption}

\def\aux@caption[#1]#2{
   \parindent 20pt \par
      {\refstepcounter\@captype \@caption{\@captype}[#1]{#2}}}

\long\def\@caption#1[#2]#3{\par\addcontentsline{\csname
  ext@#1\endcsname}{#1}{\protect\numberline{\csname
  the#1\endcsname}{\ignorespaces #2}}\begingroup
    \@parboxrestore
    \small\sl
    \@makecaption{\csname fnum@#1\endcsname}{\ignorespaces #3}\par
  \endgroup}

\def\maketitle{
 \begingroup
 \def\thefootnote{\fnsymbol{footnote}}
 \def\@makefnmark{\hbox
 to 0pt{$^{\@thefnmark}$\hss}}
 \if@twocolumn
 \twocolumn[\@maketitle]
 \else 
 \global\@topnum\z@ \@maketitle \fi
 \thispagestyle{empty}
 \setcounter{page}{0}
 \@thanks
 \endgroup
 \setcounter{footnote}{0}
 \let\maketitle\relax
 \let\@maketitle\relax
 \gdef\@thanks{}\gdef\@author{}\gdef\@title{}\let\thanks\relax}

\def\paperid#1{\gdef\@paperid{#1}}

\def\@maketitle{

 \@makepub
 \vskip 4em \begin{center}
 { \Large \bf \@title \par}
 \vskip 1.5em {\large \lineskip   .5em
 \@authoraddress
 }
 \end{center}
 \par
 \vskip 1.5em
}

\def\@makepub{{
  \centering
  \makebox[\textwidth]{
    \parbox[t]{0.25\textwidth}{\begin{flushleft}%
      {\small\@pubdate}\end{flushleft}}
    \hfil
    \parbox[t]{0.5\textwidth}{\begin{center}%
      {\small \@publabel}\end{center}}
    \hfil
   \parbox[t]{0.25\textwidth}{\begin{flushright}{\small
    \@pubnumber}\end{flushright}}
  }
}}

\gdef\@publabel{\hfil}
\gdef\@pubdate{Jan 1, 1999}
\gdef\@pubnumber{EFI-??-??}

\long\def\pubdate#1{\gdef\@pubdate{#1}}
\long\def\pubnumber#1{\gdef\@pubnumber{#1}}
\long\def\publabel#1{\gdef\@publabel{#1}}

\else
  \font \crest=uofc-shield
  \gdef\@publabel{\crest C}
\fi
\def\abstract{\if@twocolumn
\section*{ABSTRACT}
\else \small
\begin{center}
{
ABSTRACT\vspace{-.5em}\vspace{0pt}}
\end{center}
\quotation
\fi%
}

\def\endabstract{\if@twocolumn\else\endquotation\fi}

\def\pacs#1{\par %
  \bgroup
  \hsize\columnwidth \parindent0pt
  \if@twocolumn\else\leftskip=0.10753\textwidth \rightskip\leftskip\fi
 \ifdim\prevdepth=-1000pt \prevdepth0pt\fi
 \dimen0=-\prevdepth \advance\dimen0 by20pt\nointerlineskip
  \vbox to28pt{\small\vrule height\dimen0 width0pt\relax%
	PACS: #1\vfill}
  \egroup
  \if@twocolumn\vskip1pc\fi
  \newpage
}

\gdef\@author{Nobody}
\gdef\@authoraddress{}

\def\@makeauthor{
  {\def\and{\smallskip {\normalsize \rm and\smallskip}}
  {\zerospfalse \centering \large \@author}
  }
}

\def\author#1{\expandafter\def\expandafter\@authoraddress\expandafter
  {\@authoraddress %
  {
  \dimen0=-\prevdepth \advance\dimen0 by23pt
  \nointerlineskip
  \rm\centering
  \vrule height\dimen0 width0pt\relax\ignorespaces#1%
      \\[\baselineskip] 
  }%
  }%
}

\def\address#1{\expandafter\def\expandafter\@authoraddress\expandafter
  {\@authoraddress{\small\it\centering \baselineskip 1.3\baselineskip
\ignorespaces#1 \par
  }}
}

\def\thebibliography#1{\newpage
\section*{References\markboth{REFERENCES}{REFERENCES}}
\addcontentsline{toc}{section}{References}\labelsep1.0em\list
  {\arabic{enumi}.}{\settowidth\labelwidth{#1.}%
  \leftmargin\labelwidth
    \advance\leftmargin\labelsep\usecounter{enumi}}}

\let\acknowledgements=\acknowledgement

\def\msubsection{\@startsection{subsection}{2}{0.25em}%
  {4.5ex plus 1ex minus .2ex}{1.0ex plus .2ex}{\normalsize\bf}}

\def\appendix{\par
  \setcounter{section}{0}
  \setcounter{subsection}{0}
  \setcounter{equation}{0}
  \def\theequation{\thesection.\arabic{equation}}
  \def\thesection{\Alph{section}}}

\def\wider#1{\dimen1=#1 \divide \dimen1 by 2
  \advance \textwidth by #1 \advance \oddsidemargin by -\dimen1
  \advance \marginparwidth by -\dimen1  \evensidemargin\oddsidemargin
  \hsize\textwidth}
\def\taller{\advance \textheight by 50pts \vsize\textheight}
\def\shorter{\advance \textheight by -50pts \vsize\textheight}

\advance \topmargin -3\baselineskip
\headheight=\baselineskip \headsep=2\baselineskip
\def\twoup{
        \mytwocolumn
	\sloppy\flushbottom\parindent 2em
        \leftmargini 2em\leftmarginv .5em\leftmarginvi .5em
        \oddsidemargin -.5in    \evensidemargin 0in
        \columnsep .4in \footheight 0pt
        \textwidth 10in \topmargin  -.4in
        \headheight 0pt \topskip 0in
        \textheight 6.9in \footskip 30pt
        \def\@oddfoot{\hfil\thepage\hfil\addtocounter{page}{1}
                \hspace{\columnsep}\hfil\thepage\hfil}
        \let\@evenfoot\@oddfoot \def\@oddhead{} \def\@evenhead{}
}

\def\mytwocolumn{
   \global\columnwidth\textwidth
   \global\advance\columnwidth -\columnsep \global\divide\columnwidth\tw@
   \global\hsize\columnwidth \global\linewidth\columnwidth
   \global\@twocolumntrue \global\@firstcolumntrue
   \@dblfloatplacement\@ifnextchar[{\@topnewpage}{} 
}

\newdimen\slashraise \slashraise=0.33pt
\mathchardef\fslash="0236

\def\slash@char#1#2{%
   \setbox0=\hbox{$\m@th#2$}
   \dimen0=\wd0                                 
   \dimen2=-\dp0 \advance\dimen2 by \slashraise
   \setbox1=\hbox{$\m@th#1\mkern-13mu\fslash$}
	 \dimen1=\wd1               
   \ifdim\dimen0>\dimen1                        
      \rlap{\hbox to \dimen0{\hss\raise\dimen2\box1\hss}}%
      #2                          		
   \else                                        
      \rlap{\hbox to \dimen1{\hss\box0\hss}}    
      \raise\dimen2\box1                              
   \fi}                                         %

\def\slashchar#1{\mathpalette\slash@char#1}
\let\slsh=\slashchar

\def\vereq#1#2{\lower3pt\vbox{\baselineskip1.5pt \lineskip1.5pt
  \ialign{$\m@th#1\hfill##\hfil$\crcr#2\crcr\sim\crcr}}}

\def\bar{\overline}
\def\eqb{\begin{equation}}
\def\eqe{\end{equation}}

\def\hardfill#1{\vrule depth \z@ height\z@ width #1}

\def\MeV{{\rm \,MeV}}
\def\GeV{{\rm \,GeV}}

\def\etal{{\it et al.}}

\let\goesto\rightarrow

\def\tr{\mathop{\rm tr}\nolimits}

\def\trd{\mathop{\rm tr}_D\nolimits}

\def\str{\mathop{\rm str}\nolimits}
\def\sdet{\mathop{\rm sdet}\nolimits}
\def\det{\mathop{\rm det}\nolimits}

\def\mybar#1{\kern 0.8pt\overline{\kern-0.8pt#1\kern-0.8pt}\kern 0.8pt}
\def\scr#1{{\cal #1}}

\def\im{{\rm i}}
\def\dfour{\frac{\Delta}4}
\def\dmthree{\frac{-3 \Delta}4}

\def\L#1{ {#1^2 \over 16 \pi^2 f^2} \log\frac{#1^2}{\mu^2} \,}
\def\Lq{ {m_0^2/3 \over 16 \pi^2 f^2}\log\frac{M_d^2}{\mu^2}\,}
\def\LMd{ {M_d^2 \over 16 \pi^2 f^2}\log\frac{M_d^2}{\mu^2}\,}

\def\Lfs{ { m_0^2/3 \over 16 \pi^2 f^2}
     {\Delta^2 F({M_d \over \Delta})
	\over M_d^2 - \Delta^2}\,}
\def\onpif#1{{#1 \over 16 \pi^2 f^2}}

\def\logm#1{\log({#1^2\over \mu^2})}

\def\qt{{\tilde q}}
\def\qbar{{\overline q}}
\def\qtbar{{\overline\qt}}
\def\phit{{\tilde\phi}}

\def\Bt{{\tilde B}}
\def\etat{{\tilde \eta}}
\def\Aslsh{\slsh A}

\publabel{\ 
	}
\pubdate{November, 1994}
\pubnumber{IFT-94-10\\ hep-ph/9412228}
\begin{document}
\begin{titlepage}
\title{
 Quenched Chiral Corrections to Heavy-Light
 Decay Constants at Order ${\bf 1/M}$
}

\author{Michael J.~Booth%
}
\address{%
	Institute for Fundamental Theory and Department of Physics\\
	University of Florida, Gainesville, Florida 32611 \\
	{\tt booth@phys.ufl.edu}
}
\maketitle

\begin{abstract}

Quenched chiral perturbation is used to study the decay
constants for heavy-light mesons beyond the leading order in $1/M$.
The results are used to estimate the error in quenched lattice
calculations of the decay constants.
For the double ratio $R_1 = (f_{D_s}/f_D)/(f_{B_s}/f_B)$ we find
that the error is small --- conservatively 5\%, but likely smaller.
We also find that
quenching decreases the ratio $f_{D_s}/f_D$ relative to the unquenched
theory.
\end{abstract}

\pacs{12.38.Gc, 12.39.Fe, 12.39.Hg}
\end{titlepage}
\section{Introduction}


The decay constants of the heavy-light mesons ($B$ and $D$ families)
play a leading role in the physics of those mesons.  In particular,
$B\,\bar B$ mixing is proportional to $f_B^2$.  Thus, knowledge of
$f_B$ is important for estimates of mixing rate and the extraction of
the CKM angles.
%
Consider the experimental quantity
\eqb
R_2 = {(\Delta M/\Gamma)_{B_s} \over (\Delta M/\Gamma)_B}.
\eqe
To a good approximation,
\eqb
R_2 = \left|\frac{V_{ts}}{V_{td}}\right|^2
	\left(\frac{f_{B_s}}{f_B}\right)^2 \frac{B_{B_s}}{B_B}.
\eqe
However, in order for this to provide a meaningful determination of
$|V_{ts}/V_{td}|$, the ratios $f_{B_s}/f_B$ and $B_{B_s}/B_B$
must be know precisely.  They have been studied within
heavy meson chiral perturbation theory (HChPT)%
\cite{Grinstein:fDs,Goity:ChHQ,Grinstein:R1,Boyd:R1} with the
result that $B_{B_s}/B_B$ is close to unity, while $f_{B_s}/f_B$
is about 1.2.
The large deviation from unity of this latter quantity
casts doubt on the reliability of HChPT.  Because of this
Grinstein\cite{Grinstein:R1} has advocated using instead
$R_1 \times f_{D_s}/f_D$, where
\eqb
        R_1 = \frac{f_{B_s}}{f_B}/\frac{f_{D_s}}{f_D}.
\eqe
and $f_{D_s}/f_D$ is taken from experiment.

The advantage of this approach is that the ratio $R_1$ is sensitive
only to corrections which violate {\em both} heavy quark and chiral
symmetries.  From the point of view of the heavy quark effective
theory, this may be obvious, but it is nonetheless worth emphasising
the idea behind it, which is simply that the long distance physics
described by the chiral theory is insensitive to the actual mass of
the heavy quark.  Thus, one expects that while the chiral corrections
may be badly behaved, their variation with respect to $M$ (probed by
$R_1$) should be under control.
Indeed, this expectation was born-out in a recent calculation by Boyd
and Grinstein\cite{Boyd:R1} (BG) which found that $R_1$ differs from
unity by only $-5\%$.

The properties of heavy mesons have also been studied intently on the
lattice.  In particular, there have been many computations
of $f_B$ and several of $B_B$.  Thus the lattice provides
an alternate way to determine $R_1$.  But of course it is necessary
to know the error in these determinations.
One of the systematic
errors still present in most lattice calculations is that which arises
from the use of the quenched (or valence) approximation, in which
disconnected quark loops are neglected.  Quenching alters both the
short and
long-distance properties\footnote{
	Here long and short distances are defined relative to the
	QCD scale.}
of the theory.
The short-distance the effects can, in general,
be accounted for by appropriately adjusting the couplings of the
theory, but the long-distance effects are more elusive.
In a previous paper (\citen{Booth:QChHQ1}), these effects were
studied in heavy-light mesons by extending
quenched ChPT%
\cite{BernGolt:QChPT,Sharpe:QChPT1,Sharpe:QChPT2,BernGolt:PQChPT}
to include
heavy mesons at leading order in $1/M$.  It was found
that quenching has a relatively minor effect on $B_B$ and the
Isgur-Wise function $\xi(w)$, but a more pronounced effect
on $f_B$.  However, the argument in the preceding paragraph suggests
that the long-distance effects of quenching
should largely independent of the heavy quark mass.
Thus, one might expect that $R_1$ and similar quantities will be
relatively immune to the effects of quenching.

In this paper we will attempt study this supposition in more
detail.  Our tool will again be quenched ChPT.
%
%
Building
on the work of BG, we will extend this
to include $1/M$ corrections and use it to examine the
$1/M$ dependence of the quenched corrections.
By comparing these
corrections to those found in unquenched ChPT by BG, one obtains
an estimate of the error due to the use of the quenched approximation.
Of course, the conclusions drawn from this
approach are only reliable to the extent that
ChPT accurately describes the unquenched physics.
The remainder of this paper is organized as follows.
In section 2, we briefly review quenched ChPT
for heavy mesons.
In section 3, we compute loop corrections to the heavy meson
decay constants.
In section 4 these results are investigated numerically.
In section 5 we conclude.
An appendix collects some results on the one-loop counter-terms.
For the sake of brevity, we will follow the notation of BG as much
as possible and refer the reader to their paper for
omitted details.

\section{Quenched Chiral Perturbation Theory and the Inclusion of
Heavy Mesons}

Quenched chiral perturbation theory has been developed in
refs.~\citen{BernGolt:QChPT,BernGolt:PQChPT}
and the extension to heavy mesons fields
has recently been discussed in Ref.~\citen{Booth:QChHQ1}.
Consequently
we will restrict the presentation here to only that which is
necessary to fix notation.

On way to implement quenching\cite{Morel:QLogs} is to introduce
 bosonic ``ghost''
quarks to cancel the functional determinant which arises
from the integral over the fermion fields.  The inclusion of these
extra particles enlarges the symmetry of the theory.
Thus,
quenched ChPT is obtained from ordinary
unquenched ChPT by enlarging the symmetry group $SU(3)_L\times SU(3)_R$
to the semi-direct product
$(SU(3|3)_L\times SU(3|3)_R){\bigcirc\kern -0.75em s\;}U(1)$.
Elements of the graded symmetry group are represented by supermatrices
(in block form)
\eqb
U=\left(\matrix{A&B\cr C&D\cr}\right),
\eqe
where $A$ and $D$ are matrices composed of even (commuting) elements
and $B$ and $C$ are composed of odd (anti-commuting) elements.
Group invariants are formed using the
super trace $str$ and super determinant $sdet$, defined as
\begin{eqnarray}
\str(U) &=& \tr(A)-\tr(D), \\
\sdet(U) &=& \exp(\str\log{(U)})=\det(A-B D^{-1} C)/\det(D).
\end{eqnarray}
To accommodate the larger symmetry, the meson matrix is
extended to a supermatrix:
\eqb
\Phi = \left(\matrix{\phi&\chi^\dagger\cr
\chi&\phit\cr}\right),
\eqe
where $\chi^\dagger \sim \qt \qbar$, $\chi \sim q \qtbar$ and
$\phit \sim \qt\qtbar$ and $\phi$ is ordinary meson matrix
\eqb
\phi =
\pmatrix{\frac 1{\sqrt 2}\pi^0 + \frac 1{\sqrt 6}\eta &
\pi^+ & K^+ \cr
\pi^- & -\frac 1{\sqrt 2} \pi^0 + \frac 1{\sqrt 6}\eta &
K^0 \cr
K^- & {\mybar K}^0 & -\sqrt{\frac 2 3} \eta \cr}.
\eqe
Note that $\chi$ and $\chi^\dagger$
are fermionic fields, while $\phi$ and $\phit$ are bosonic.

The Lagrangian of quenched ChPT is given by
\eqb
\label{eqn:LBG}
\scr{L}_{Q\chi} = {f^2\over 8} \left[
        \str(\partial_\mu\Sigma\partial^\mu\Sigma^\dagger)
        +4\mu_0\,\str(\scr{M}_{+}) \right]
    + \frac{\alpha_0}2 \partial_\mu\Phi_0\partial^\mu\Phi_0
    - \frac{m_{0}^2}2 \Phi_{0}^2.
\eqe
Here $\Sigma = \xi^2$, $\xi = e^{i\Phi(x) / f}$
(the normalization is such that $f_\pi = 128 \MeV$)
while
\begin{eqnarray}
\Phi_0 & = &{1\over\sqrt3}\, \str \Phi =
	{1\over \sqrt2}(\eta' - \etat'), \\
\scr M &=& \left(\matrix{M&0\cr 0&M\cr}\right), \\
M &=& \pmatrix{m_u &&\cr & m_d &\cr && m_s \cr},
\end{eqnarray}
and $\scr M_\pm =
	\frac 12 (\xi^\dagger \scr M \xi^\dagger \pm \xi \scr M \xi)$.
The chief
difference between the quenched and unquenched theories is
the presence of the terms involving $\Phi_0$.  In the unquenched
theory they can be neglected because they describe the dynamics
of the $\eta'$ meson, which decouples from the theory.  But quenching
prevents the $\eta'$ from becoming heavy and decoupling, so these
terms must be retained in quenched ChPT.

The propagators that are derived from this Lagrangian are the ordinary
ones, except for the flavor-neutral mesons, where the non-decoupling
of $\Phi_0$ leads to a curious double-pole structure.
It is convenient to adopt a basis $U_a$ for the those
mesons corresponding to
$u\mybar u, d\mybar d$ and so on, including the ghost quark
counterparts.
Then the propagator takes the form
\eqb
G_{ij} = {\delta_{ij} \epsilon_i \over p^2 - M_i^2}
+ {(-\alpha_0 p^2 + m_0^2)/3 \over (p^2 - M_i^2)(p^2 - M_j^2)}
\eqe
where $\epsilon = (1,1,1,-1,-1,-1)$ and $M^2_i = 2\mu_0 m_i$.  It is
conventional to
treat the second term in the propagator as a new vertex,
the so-called hairpin, with the rule that it can be inserted only once
on a given line.  Large $N_c$ arguments suggest that
the kinetic coupling $\alpha_0$ is small; this is supported by an
analysis of $\eta-\eta'$ mixing.  Consequently, we 
follow the usual practice and set it to zero in the sequel.

In heavy quark effective theory\cite{Wise:ChHQ,Burdman:ChHQ,Yan:ChHQ}
the $B$ and $B^*$ fields
 (for convenience we refer to all heavy mesons as B's)
are grouped into
the $4 \times 4$ matrix $H_a$ which conveniently encodes the
heavy quark spin symmetry:
\begin{eqnarray}
H_a &=& \frac12(1+\slsh v)[\bar B^{*\mu}_a\gamma_\mu -
	\bar B_a\gamma_5],\\
	\bar H_a &=&\gamma^0 H_a^\dagger \gamma^0\,.
\end{eqnarray}
Here $v^\mu$ is the four-velocity of the heavy meson,
the index $a$ runs over the light quark flavors, $u$, $d$, $s$ and
the subscript ``D'' indicates that the trace is taken only over
Dirac indices.
Henceforth we will
drop explicit reference to the heavy meson velocity.
The quenched
heavy mesons can be incorporated into this framework by adding to $H$
extra fields $\Bt$ and $\Bt^*$ derived from the heavy fields $B$ and
$B^*$ by replacing the light quark with a ghost quark, so that $a$
also runs over the ghost flavors.

BG (see also Ref.~\citen{KuriKita:HQ})
have given the chiral heavy quark Lagrangian to
order $1/M$ (throughout the paper $M$ will refer to
the spin-averaged meson mass,  $M = \frac14 (M_B + 3 M_{B^*})$).
To formulate the quenched version one must also
include in the Lagrangian vertices which couple $\Phi_0$ to $H$.
Symmetry
requires that this coupling occur
through $\str(A_\mu)$, which no longer vanishes.
Adding these vertices and some extra counter-terms, but including
only those terms which actually contribute at $O(1/M)$, one finds
\begin{eqnarray}
\label{eqn:ChHQL}
{\cal L}&=&
  -(1 + \frac{\epsilon_1}M )\trd\left[
            \mybar H_a i v\cdot D_{ba} H_b\right]
	+ \frac{\epsilon_2}M
	  \trd\left[
         \mybar H_a \sigma^{\mu \nu} v\cdot D_{ba} H_b\sigma_{\mu \nu}
	  \right] \nonumber \\
  &&\mbox{} + ( g + \frac{g_1}M) \trd\left[\overline
             H_a H_b \,\Aslsh_{ba}\gamma_5\right]
     + \frac{g_2}{M} \trd\left[\overline H_a  \Aslsh_{ba}
            \gamma_5 H_b \right]  \nonumber \\
  &&\mbox{} + ( \gamma + \frac{\gamma_1}M) \trd\left[\overline
             H_a H_a \gamma_\mu\gamma_5\right] \str(A^\mu)
        + \frac{\gamma_2}{M} \trd\left[\overline H_a  \gamma_\mu
            \gamma_5 H_a \right]\str(A^\mu)   \nonumber \\
  &&\mbox{} +
	\lambda_1 \trd\left[\overline H_a H_b \right] (\scr{M}_{+})_{ba}
		+ \frac{\lambda_2}M \trd\left[\overline H_a
             		\sigma^{\mu \nu}H_a  \sigma_{\mu \nu}\right]
  + {\lambda_3\over M}
     \trd\left[\overline H_a \sigma^{\mu\nu}H_b\sigma_{\mu\nu} \right]
	(\scr{M}_{+})_{ba}
   \nonumber\\
  &&\mbox{} + (k_{10} + {k_{11}\over M}) \trd\left[
            \mybar H_a i v\cdot D_{bc} H_b\right](\scr{M}_+)_{ca}
   +\frac{k_3}M \trd\left[
       \mybar H_a \sigma^{\mu \nu} v\cdot D_{bc} H_b\sigma_{\mu \nu}
      \right](\scr{M}_+)_{ca}.
\end{eqnarray}
Light mesons enter this Lagrangian through
the quantities:
\begin{eqnarray}
D_\mu &=& \partial_\mu + V_\mu, \nonumber \\
V_\mu &=& \frac 12\left(\xi \partial_\mu \xi^\dagger
+ \xi^\dagger \partial_\mu \xi\right), \\
A_\mu &=& \frac i2\left(\xi \partial_\mu \xi^\dagger
-\xi^\dagger \partial_\mu \xi\right) =
	  - {1\over f}\partial_\mu \Phi + \scr{O}(\Phi^3).
\end{eqnarray}
The couplings $\lambda_i$ shift the masses of the heavy mesons: the
$B^*$-$B$ mass splitting $\Delta$
is given by $\Delta = -8\lambda_2/M$, while
$\lambda_1$ produces an $SU(3)$ violating mass shift
$\delta_q = 2\lambda_1 m_q$
and $\lambda_3$ violates both symmetries
(though to the order we are working it
only contributes at tree level).
Taking these shifts into account the $B_q$ and $B_q^*$
propagators become
${i\over 2(v\cdot k + \frac34\Delta-\delta_q)}$
and
${-i(g_{\mu\nu}-v_\mu v_\nu)\over
	2(v\cdot k - \frac14\Delta-\delta_q)}$,
respectively.
The propagators of the ghost mesons are the same as
their real counterparts.
It is convenient to follow BG in
adopting a shorthand notation for the Lagrangian
(with the
understanding that the propagators are the shifted one):
\eqb
\scr L = -\trd\left[\overline H_a\im
 v\cdot D_{ba} H_b\right] + \tilde g_{\mybar H H} \,\trd\left[
 \overline H_a H_b\,\Aslsh_{ba}\gamma_5\right]\, +
 {\tilde \gamma_{\mybar H H} }\,\trd\left[
 \overline H_a H_a\,\gamma_\mu\gamma_5\right] \str(A^\mu),
\eqe
where
\eqb
\tilde g = \cases{ {\tilde g}_{B^*} =
      g + \frac1M (g_1 + g_2) & for $B^* B^*$ coupling,\cr
	\vphantom{|}\cr
\tilde g_{B^{\phantom{*}}} =
      g + \frac1M (g_1 - g_2) &for $B^* B^{\phantom{*}}$ coupling,\cr}
\eqe
and similarly for $\tilde\gamma$.  In the sequel it will be
useful to introduce two more couplings,
\eqb
{\hat g}_{B^*} = g + {(g_1+g_2/3)\over M}, \qquad
	{\hat \gamma}_{B^*} = \gamma + {(\gamma_1+\gamma_2/3)\over M}.
\eqe
All of these couplings should be regarded as shorthand.
That is, when performing arithmetic with these
couplings, only terms to $O(1/M)$ are to be retained.
%
Finally, we record the expression for the weak current%
\footnote{Here and in the Lagrangian \ref{eqn:ChHQL} several
	terms proportional to $\str(\scr{M}_+)$ have been since
	they do not contribute, but their presence can be inferred
	from the gaps in the numbering of the coefficients.
}
(again with counter-terms)
\begin{eqnarray}
\label{eqn:currentfull}
J_a^\mu &=&  i\alpha (1 + \frac{\rho_1}M  )
     \trd[\Gamma^\mu H_b\xi^\dagger_{ba}]
  +  i\alpha \frac{\rho_2}{M}
      \trd[\gamma^\sigma \Gamma^\mu \gamma_\sigma H_b\xi^\dagger_{ba}]
  + \frac{i\alpha}{2 M} \trd\left[ [\Gamma^\mu,\gamma_\sigma]
            \im D^\sigma_{cb} H_c\xi^\dagger_{ba} \right] \nonumber\\
&&\mbox{}+ i \alpha (\kappa_{10} + \frac{\kappa_{11}}M  )
      \trd[\Gamma^\mu H_c\xi^\dagger_{cb}] (\scr{M}_+)_{ba}
 + i \alpha \frac{\kappa_3}{M}
  \trd[\gamma^\sigma \Gamma^\mu
	\gamma_\sigma H_c\xi^\dagger_{cb}](\scr{M}_+)_{ba},
%
%
%
\end{eqnarray}
where $\Gamma^\mu = \gamma^\mu L = \gamma^\mu (1-\gamma_5)/2$.

\section{Loop Corrections}
In ChPT, loop corrections generate terms which are non-analytic in
the mass parameters of the theory.  These terms are uniquely
predictions because they cannot be canceled by higher-order
counter-terms.  The loops also generate divergences and finite terms
which are analytic; these are absorbed into the counter-terms.
In principle, they can be determined by fitting to experimental
data,  but due to the paucity of data in the heavy-meson sector,
we will at times be forced to simply neglect them and they will
not be shown in our formulas.  We will
later discuss the uncertainty arising from this approximation,
but fortunately the error in $R_1$, which is our main interest,
should be free of counter-terms.

As noted in Ref.~\citen{Booth:QChHQ1}, the loop structure of the
quenched theory is rather odd:  the theory splits into
three copies of a one-flavor theory, distinguished only by the
light quark mass.  Because of this, all results
for $B_d$ or $B_u$ (henceforth we will refer to these simply as $B$)
mesons apply for $B_s$ mesons after an appropriate relabeling.

The loop integrals encountered are either identical to
the $I_1$ and $J$ of BG, or may be
obtained from them by differentiation with respect to $m^2$,
which will be denoted with a prime.  More explicitly,
we have $J(m, \Delta) = \Delta J_1(m, \Delta)$ and
\begin{eqnarray}
    I_1(m)&=&m^2\ln(m^2/\mu^2)\,,\\
    J_1(m,\Delta)&=&(-m^2+\frac23\Delta^2)\ln(m^2/\mu^2)+\frac43
    (\Delta^2-m^2)F(m/\Delta)\,,\\
   \vphantom{+}\nonumber\\
    F(x)&=&\cases{
    \sqrt{1-x^2}\;\tanh^{-1}\sqrt{1-x^2}\,,\,x\le1,\cr
    \vphantom{|}\cr
    -\sqrt{x^2-1}\;\tan^{-1}\sqrt{x^2-1}\,,\,x\ge1.\cr}
\end{eqnarray}

\begin{figure}
\ifIncludeFigs
  \centerline{\epsffile{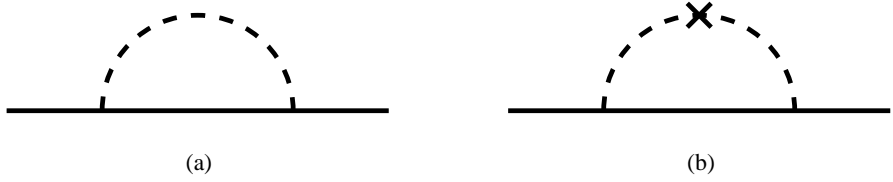}}
\fi
\caption{\label{fig:se}
	The diagrams which contribute to the heavy meson self energy.
	Solid lines represent heavy mesons ($B$, $B^*$), dashed lines
	represent light mesons and the cross represents an insertion
	of the ``hairpin'' vertex.
}
\end{figure}

The graphs which contribute to the self-energy are
shown in Fig.~\ref{fig:se}.
%
In the diagram of Fig.~\ref{fig:se}a,
the ghost mesons will cancel the contribution from
the real mesons unless one of the vertices involves the singlet field.
Combining this with the contribution of the hairpin
vertex Fig.~\ref{fig:se}b,
we obtain (recall that we are taking $\alpha_0$ to vanish)
%
\eqb
\label{eqn:singletloop}
i\,\Sigma_B(v\cdot k) =
{6 i \over 16 \pi^2 f^2}\left[
  2\tilde g_B \tilde\gamma_B
	J(M_d, \dfour + \delta_d - v \cdot k )
 + \tilde g_B^2 {m_0^2 \over 3}
	J'(M_d, \dfour + \delta_d - v \cdot k )
 + \,\ldots\,
  \right].
\eqe
The terms not shown are analytic in $M_d$ and are absorbed in the
definition of the counter-terms.
For the $B^*$, we similarly obtain
\begin{eqnarray}
i\,\Sigma^{\mu\nu}_{B^*}(v\cdot k)\!\! & = &\!\!
   {-2 i g^{\mu\nu} \over 16 \pi^2 f^2}\left[
   \tilde g_B^2 {m_0^2\over3}
		J'(M_d, \dmthree + \delta_d - v \cdot k)
	+ 2\tilde g_{B^*}^2 {m_0^2\over3}
		J'(M_d, \dfour + \delta_d - v \cdot k )
	 \right. \nonumber\\
  &&+ \left.
   2\tilde g_B \tilde\gamma_B
	J(M_d, \dmthree + \delta_d - v \cdot k )
   +4\tilde g_{B^*}\tilde\gamma_{B^*}
	J(M_d, \dfour + \delta_d - v \cdot k ) + \,\ldots
        \right] ,
\end{eqnarray}
where we have neglected a term proportional to $v^\mu v^\nu$.
Note that the $\delta_d$ dependence will vanish when the on-shell
conditions are invoked.

\subsection{Wavefunction Renormalization and Decay Constants}

The wavefunction renormalization constants are obtained by
differentiating the self-energy with respect to $2v\cdot k$
and evaluating on-shell.  BG made no assumptions
about the magnitude of ${m_\pi\over \Delta}$ and thus retained
the function $F$.
In lattice simulations,
where both $m_\pi$ and $\Delta$ are adjustable parameters, similar
restraint is warranted and so we will also retain $F$.
In terms of the shorthand couplings introduced earlier, the
renormalization constants can be written compactly as
\begin{eqnarray}
  Z_{B} &=& 1
      +  3 {\tilde g_B}^2 \Lq
	+ 6 {\tilde g_B} {\tilde \gamma_B} \LMd \nonumber\\
	&&\mbox{} - 6 g^2 \Lfs
	- 24 g \gamma \onpif{\Delta^2}F({M_d\over \Delta}), \\
%
  Z_{B^0} &=& 1
  + 3 {\hat g_{B^*}}^2 \Lq + 6 {\hat g_{B^*}} {\hat \gamma_{B^*}}\LMd
  \nonumber\\
	&&\mbox{} - 2 g^2 \Lfs
	- 8 g \gamma \onpif{\Delta^2}F({M_d\over \Delta}).
\end{eqnarray}
The above expressions have been simplified by discarding terms that are
formally of higher order in the chiral and large mass expansions.
However, it is important to
retain those terms which are needed to make
the chiral limit consistent.

\begin{figure}
\ifIncludeFigs
  \centerline{\epsffile{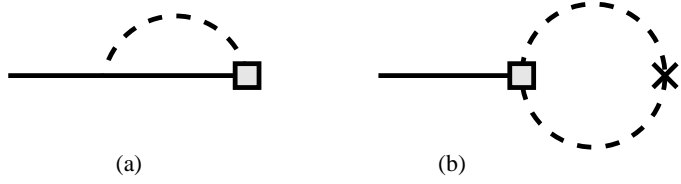}}
\fi
\caption{\label{fig:vertex}
	The tadpole correction to the weak current vertex.  The
	box represents an insertion of the weak current.
}
\end{figure}

Loop corrections to the left-handed current vertex arise from the
diagrams of Fig.~\ref{fig:vertex}.
As was shown by BG, the diagrams of
Fig.~\ref{fig:vertex}a do not contribute at $O(1/M)$.
The remaining
tadpole graph Fig.~\ref{fig:vertex}b yields for the $B$
\eqb
{\im \alpha v^\mu \over 2 }
   (1 + {\rho_1 + 2 \rho_2 \over M})\Lq
\eqe
and for the $B^*$,
\eqb
{-\im \alpha \epsilon^\mu \over 2}
   (1 + \frac{\rho_1 - 2 \rho_2}{M})\Lq.
\eqe
The final results for the decay constants
are found by combining
the wavefunction and vertex corrections:
\begin{eqnarray}
\sqrt{M_{B}}f_{B} &=&
   \alpha \left( 1 + {\rho_1 + 2 \rho_2 \over M} \right)
   \left(
      1 - \frac12 (1 + 3 {\tilde g_B}^2)\Lq
      - 3 {\tilde g}_B {\tilde \gamma}_B \LMd
   \right. \nonumber \\
   &&\left. \mbox{} + 3 g^2 \Lfs +
   12 g \gamma \onpif{\Delta^2}F({M_d\over \Delta}) \right), \\
\frac1{\sqrt{M_{{B}^*}}}f_{{B}^*}
    &=& \alpha \left( 1 + \frac{\rho_1 -2\rho_2}{M}\right)
    \left(
     1  - \frac12 (1 + 3{\hat g_{B^*}}^2) \Lq
       -3 {\hat g_{B^*}} {\hat \gamma_{B^*}} \LMd
    \right. \nonumber \\
    && \left.\mbox{} +  g^2 \Lfs
      + 4 g \gamma \onpif{\Delta^2}F({M_d\over \Delta}) \right).
\end{eqnarray}
At this point, let us pause to point out that the nontrivial $1/M$
dependence enters only only through the wavefunction renormalization
graphs.  The tadpole corrections are only multiplicative --- they
do not see the heavy meson mass.
It follows that if the $B^* B\pi$
couplings are weak, the light and heavy quark dependencies of
the decay constants will decouple from each other.

To study the size of $1/M$ corrections it is useful to consider
the ratio $U(M) = M f_{B}/f_{B^*}$, which is one in the
infinite mass limit.
In the quenched theory we find

\begin{eqnarray}
  U(M)  &=& 1 + { 4 \rho_2 \over M}
	+ {4 g g_2 \over M} \Lq
	+ {4(g \gamma_2 + \gamma g_2) \over M} \LMd \nonumber\\
	&&\mbox{} + 2 g^2 \Lfs
         + 8 g \gamma \onpif{\Delta^2}F({M_d\over \Delta}),
\end{eqnarray}
while in the full theory
\begin{eqnarray}
\label{eqn:Ufull}
  U(M)  &=& 1 + { 4 \rho_2 \over M}
	+ \frac{11}{9} {4 g g_2 \over M} \L{m_K}
	+ 8g^2 \onpif{\Delta \delta} \logm{m_K} \nonumber \\
	&&\mbox{} + 6g^2 (\Delta+\delta)^2 F({m_K\over \Delta+\delta})
	- 2g^2 (\Delta-\delta)^2 F({m_K\over \Delta-\delta}),
\end{eqnarray}
which is easily derived from the
work of BG.  Here $\delta = \delta_s -\delta_d$ and in order to
keep the expression compact we have followed BG in neglecting
the pion contribution and approximation $m_\eta^2 = \frac43 m_K^2$,
but the exact expression is used for numeric work.

\section{Numeric Results}
When studying the error due to quenching, it is
interesting to compare our quenched predictions not just with the
predictions of BG, but also with the predictions of
an unquenched two-flavor theory with degenerate quark masses.
Such a theory should describe most unquenched lattice
simulations%
\footnote{A more accurate treatment would take into account
	 the difference between the valence and sea quark masses by
	 using partially quenched ChPT\cite{BernGolt:PQChPT}
}
and allows one to distinguish between those effects which are
due to quenching and those which are due to the simplified flavor
structure of the lattice simulations.
Results for the two flavor theory can be extracted from
the work of BG by taking into account the $SU(3)$ coefficients.
In the quenched and two-flavor theories, $f_{B_s}$ is
obtained from $f_B$ by substituting
$M_s = \sqrt{2 m_K^2 - m_\pi^2} = 680 \MeV$ for $M_d$.
%
For convenience, 
the mass of the light mesons in the quenched and two-flavor
theories will always be referred to as $M_d$.  Since lattice
simulations typically have $M_d > 400\MeV$, it will be useful
to study the $M_d$ dependence of our results.

Before proceeding, it is necessary to specify the
various couplings which enter the Lagrangian (\ref{eqn:ChHQL}).
Unfortunately,
little information is available.
By taking into account chiral loop
corrections to strong and radiative $D^*$ decays,
one obtains the constraint\cite{Amundson:g,Cheng:g}
$0.3  < g + (g_1-g_2)/M_{D} < 0.7$, with the central
value being slightly less than $0.5$.
We will choose
$g = 0.5$ and $g_1 = g_2$ as our canonical values.
QCD sum rules\cite{QCDSR:g1,QCDSR:g2,QCDSR:g3,QCDSR:g4}
and relativistic quark models\cite{RQM:g1,RQM:g2}
favor a smaller value of g,
$ g \sim 1/3$.
They also indicate that $g_1 - g_2$ is small,
on the order of $100 \MeV$,
which is roughly consistent with our choice.
When necessary, we will assume that both $g_1$ and $g_2$ are of this
order.
This choice is supported by a recent study\cite{QCDSR:g_2} which found
$g_2 \simeq 0.3 g \GeV$.
A lattice study of heavy meson decay constants by
Bernard {\it et al.}\cite{BLS:fB}
found $\rho_1 + 2 \rho_2 = -1.14 \GeV$,
while a similar study by the UKQCD collaboration\cite{UKQCD:fB}
found $\rho_1 + 2 \rho_2 = -0.8 \GeV$.
Because the latter group also studied $f_{D^*}$ it possible
to extract $\rho_2$ from their data;  we find $\rho_2 \approx -0.1
\GeV$, which is consistent with other
determinations\cite{Sommer:Lectures}.
The remaining couplings $\gamma$ and $\gamma_i$
are unconstrained.  Relative to $g$ and $g_i$, they
are non-leading in $1/N_c$, so they are likely to be small
(in the nucleon system, the singlet coupling $\gamma$
is indeed small\cite{Hatsuda:singlet}).
In view of this, we will take $\gamma = \gamma_i = 0$, but it
should be kept in mind that this may be an incorrect assumption.
The proper choice of $m_0$ was discussed in Ref.~\citen{Booth:QChHQ1};
here we choose $m_0 = 750\MeV$.

We begin by considering the ratio $R_1$.  We first observe that
$R_1$ has only one counter-term,
whose contribution is proportional to
$(\kappa_{11}+2\kappa_3)(m_s-m_d)(1/M_D-1/M_B)$.
Since we expect from the nature of the chiral expansion
that the counter-terms
will be about the same in the quenched and full theories%
\footnote{
  We do not expect them to be exactly the same because
  the two theories have different divergences,
  but they should be close if the lattice is doing a good job
  of describing QCD},
it follows that the {\em error} in $R_1$ will only weakly depend
on the counter-term.
With this in mind, we neglect it and summarize
the results in Table~\ref{tab:R1}.%
\footnote{
  The fact that our result in the three flavor case
  agrees with that
  of BG 
  is something of a coincidence,
  because we have chosen to use $f = f_\pi$
  while they use $f = f_K$ and we have also
  retained the pion contribution.}
%
\begin{table}[tb]
\caption{\label{tab:R1}
	Quenched and unquenched results for $R_1$.
}
\begin{center}
\begin{tabular}{|ll|}
\hline
Theory & $R_1-1$
	\\
\hline
quenched & $
  -0.21 g^2 + 0.20 \GeV^{-1} g(g_1-g_2)$ \\
	& $\phantom{-0.18 g^2} + 0.23 g \gamma
    - 0.097\GeV^{-1} (g(\gamma_1-\gamma_2) + \gamma(g_1-g_2)) $
\\
$N_f = 2$ & $
    \phantom{-}0.17 g^2 - 0.15\GeV^{-1}g(g_1-g_2) $
\\
$N_f = 3$ & $
   -0.10 g^2 - 0.14 \GeV^{-1} g(g_1-g_2) $
\\
\hline
\end{tabular}
\end{center}
\end{table}
%
When the reference values are chosen, $R_1-1$ is
%
$-0.026$ in the full theory,
$-0.053$ in the quenched theory, but
$+0.043$ in the two flavor theory.
%
%
%
%
%
In order to explore the dependence of these results
on
the couplings, we show in Table~\ref{tab:R1num}
the results for several different choices of $g$ and $g_i$.
The first set corresponds to the parameters chosen
by BG, the second set is our canonical one, and the next
two explore the dependence on $g_1-g_2$.
\begin{table}[tb]
\caption{\label{tab:R1num}
	Numerical results for $R_1$ for various models and couplings.
	The couplings are chosen as follows:
	Set 1 (BG): $ g = 0.75$, $g_1 = g_2$;
	Set 2: 	    $ g = 0.5$, $g_1 = g_2$;
	Set 3: 	    $ g = 0.5$, $g_1 - g_2 = 0.2 \GeV$;
	Set 4:	    $ g = 0.5$, $g_1 - g_2 = -0.2 \GeV$.
}
\begin{center}
\begin{tabular}{|lllll|}
\hline
Theory &\multicolumn{4}{c|}{$R_1-1$}
	\\
& Set 1& Set 2& Set 3& Set 4
	\\
\hline
quenched &
	$ -0.11 $&	$-0.052$&       $-0.032$&	$-0.073$
\\
$N_f = 2$ &
	$ +0.087$&	$+0.043$&       $+0.028$&	$+0.058$
\\
$N_f = 3$ &
	$-0.052$&	$-0.025$&       $-0.040$&	$-0.012$
\\
\hline
\end{tabular}
\end{center}
\end{table}
It can be seen that the error (which is the difference between
the full and quenched results) is no more than $6\%$ and not half that
for the favored choice of couplings.
Thus as hoped
quenching has little effect on $R_1$ --- the large tadpole corrections
found in Ref.~(\citen{Booth:QChHQ1}) cancel in the ratio and the
remaining corrections are small.
Interestingly, the results
in the two-flavor case suggest that dynamical simulations will
do worse than the quenched simulations.

\begin{figure}[htbp]
\ifIncludeFigs
  \centerline{\epsfxsize=0.45\textwidth\epsffile{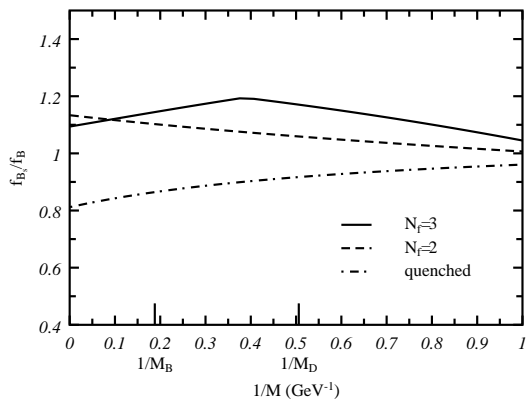}}
\fi
\caption{\label{fig:R0}
	Plot of the ratio $f_{B_s}/f_B$ in the various theories.
}
\end{figure}

The results for $R_1$ can be understood by looking at the
ratio $f_{B_s}/f_B$ as a function of $M$.
Figure~\ref{fig:R0} compares the various theories, neglecting
counter-terms.
The most striking feature is
the ``kink'' in the three-flavor
case\footnote{
  BG presumably did not comment on this kink because their work
  only required the ratio at the two physical values of $M$.
},
which occurs when $\Delta = \delta_s$,
causing the denominator in the argument
of one of the functions $F$ to vanish
  (c.f.\ the expression for $U(M)$ in Eq. (\ref{eqn:Ufull})).
Physically, this corresponds
to the case when the $B_s$ and the $B^*$ are degenerate,
leading to an enhancement of $f_{B_s}$.  Without this kink,
the ratio would be a slowly increasing function of $M$, just as
it is in the two-flavor theory.  In a sense, then, the small
error in $R_1$ is a fortunate coicidence.
While the details of the curves in Fig.~\ref{fig:R0}
depend on our choice of
$g$ and $g_i$, the shapes
are generic.
As $g$ increases,
or $g_1 - g_2$ becomes negative, the magnitudes of the slopes
increases, so that the curves draw together more at small $M$
without changing their shape or relative positions.
In particular, there is always a large gap between the quenched
and full predictions.
%
Thus a general conclusion of this analysis is that
quenching decreases the ratio $f_{D_s}/f_D$.
In addition, the different $M$ dependence
predicted by the three theories makes this ratio a test of the chiral
description
of the decay constants.
%

\begin{figure}[tbh]
\ifIncludeFigs
  \centerline{\epsfxsize=0.45\textwidth\epsffile{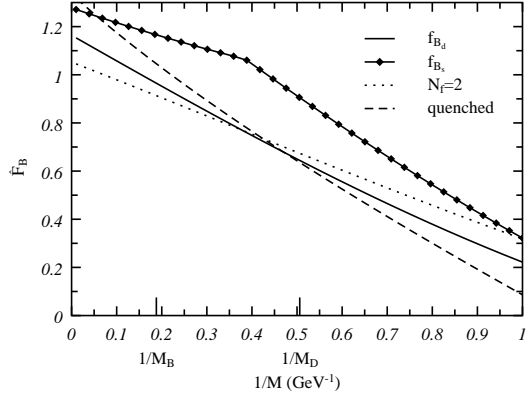}}
\fi
\caption{\label{fig:fB}
	Plot of $\hat F = \protect\sqrt{M} f_B/\alpha$
	in the various theories, with $M_d = m_\pi$.
}
\end{figure}

\begin{figure}[tbh]
\ifIncludeFigs
  \centerline{\epsfxsize=0.45\textwidth\epsffile{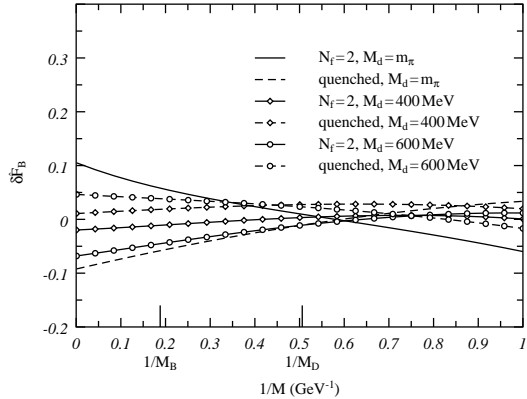}}
\fi
\caption{\label{fig:fBE}
	Plot of the differences
	$ \delta \hat F_B =
		\hat F_B({\rm quenched}) - \hat F_B({\rm full})$
	and
	$ \delta \hat F_B =
		\hat F_B(N_f=2) - \hat F_B({\rm full})$
	for various values of $M_d$.
}
\end{figure}

We next consider $f_B$ directly.  This is primarily for illustrative
purposes since the dynamical theories have an extra counter-term
proportional to $\tr(M_+)$, making direct comparison to the
quenched theory difficult; here we will simply neglect the counter-terms.
Figure~\ref{fig:fB} then shows
both $\hat F_B = \sqrt{M} f_B/\alpha$ and $\hat F_{B_s}$
together with the quenched and two flavor
predictions.
%
In the $M \goesto \infty$
limit the results are consistent with those found in
Ref.~\citen{Booth:QChHQ1}.
%
%
The behavior of $f_B$ is dominated by the tree-level term
$(\rho_1+2\rho_2)/M$.
In Fig.~\ref{fig:fBE},
we study the $M_d$ dependence of the error in $f_B$,
comparing both the quenched and two flavor theories to the
true theory.
In the error, the tree-level terms cancel, making it easier
to see the non-trivial $M$ dependence.
One sees that the $M$ and $m_q$ dependence are closely intertwined, as
the both the quenched and two-flavor errors switch sign between
$M_d = m_\pi$ and $M_d = 400\MeV$.  Even at its largest,
 the quenched error
is no more than $7\%$ in the region between $M = M_B$ and $M = M_D$.
Also, observe that except for the smallest value of $M_d$,
the error tends to decrease as $M$ decreases.

\begin{figure}[tbhp]
\ifIncludeFigs
  \centerline{\epsfxsize=0.45\textwidth\epsffile{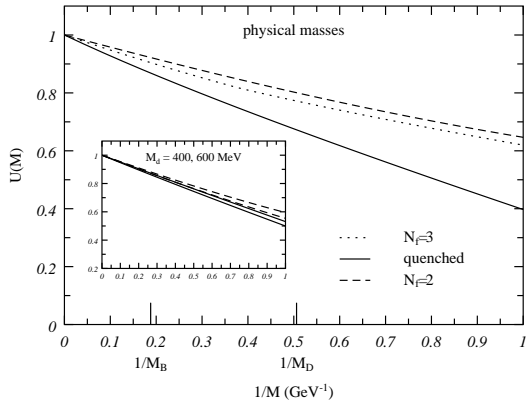}}
\fi
\caption{\label{fig:U2}
	Plot of the ratio $U(M)= M f_B/f_{B^*}$.  The inset
	graph shows the $M_d$ dependence of the quenched and
	two-flavor theories.  As $M_d$ increases, the curves
	of two theories approach each other,
	overlapping near $M_d = M_s$ and then moving apart again.
}
\end{figure}

Finally, we examine the $1/M$ dependence of $f_B$ through the
quantity $U(M)$. Once again there is the problem that the dynamical
theories have one more counter-term than the quenched theory.
This could be eliminated by considering the ratio
$U_s(M)/U_d(M)$, but we will again simply neglect the counter-terms.
In Fig.~\ref{fig:U2} we show $U$ for various values of the light
quark mass.
%
For $M_d = m_\pi$, quenching has a pronounced effect.
But for larger values of $M_d$, there is almost no difference
between the quenched and two-flavor theories.  Thus lattice
simulations are unlikely to observe any effect of quenching
in $U(M)$.  Note that since all the
theories must agree at $M=\infty$, the quenching error can only
increase as $M$ decreases.



To conclude our discussion of numeric results,
let us comment on the uncertainty arising from the
neglect of the counter-terms.  In ChPT for the light mesons, the
counter-terms are small, close to their natural size of $1/(4\pi)^2$
\cite{EGPR:Resonances}.
Goity\cite{Goity:ChHQ} has observed
that choosing $\mu = 1.5 \GeV$ effectively mimics
the inclusion of these analytic terms.
While we do not know whether
this works for the heavy mesons, it certainly does
generate terms of the natural size.
  However, it cannot be completely correct for the quenched
  case because of the different behavior of the
  quenched logarithms under scaling.
Nonetheless, it serves as
a useful way to add counter-terms of the expected size.%
\footnote{
  A more ambitious approach would be to assume that resonances
  saturate the counter-terms, as they do in the light-meson
  sector\cite{EGPR:Resonances}.
}
Doing this, 
we find that in all the theories $R_1$ is almost independent
of $\mu$, changing by only a few percent.  $U(M)$ also
changes very little, again only a few percent.  $f_B$ and
the ratio $f_{B_s}/f_B$, however, are more strongly effected:
typical shifts are $10-15\%$.  However, the general features
of $f_{B_s}/f_B$ that were observed in
Fig.~\ref{fig:R0} are unchanged.

\section{Conclusions}

By generalizing quenched ChPT to include heavy mesons and including
$1/M$ corrections following Boyd and Grinstein\cite{Boyd:R1},
we have studied the $1/M$ dependence of the quenching errors
in lattice simulations.  We found that unknown counter-terms
prevented us from making definitive statements about the behavior
of $f_B$, but it was possible to reduce this uncertainty by
taking appropriate ratios, such as $R_1$.
Although the long-distance effects of quenching
are not independent of the heavy mass $M$, they
were fairly insensitive to it.  As a result, the
error in $R_1$ was found to be small, only a few percent.
We saw that $f_{B_s}/f_B$ is smaller in the quenched theory,
a result that was observed earlier in the $M \rightarrow \infty$
limit\cite{Booth:QChHQ1}.
We also observed an effect not directly related to quenching, but
nonetheless missing in lattice simulations, namely the
enhancement of $f_{B_s}$ when $B_s$ and $B^*$ are nearly degenerate.
This enhancement had a strong impact on $R_1$ in the full theory.
In future work it would be desirable to
determine the counter-terms in some way, either from models or
directly from lattice data.

\acknowledgements
We would like to think the University of Chicago theory group for
its hospitality while this work was initiated.
This work was supported in part by DOE grant DE-FG05-86ER-40272.

\appendix
\section{Renormalized Couplings}
Within the context of dimensional regularization,
the singularities of the effective Lagrangian are conventionally
described in terms of the parameter $L(\mu)$ which contains
the singularity at $D=4$ ($\epsilon = 2-D/2$,
$1/\hat\epsilon =  1/\epsilon + \log 4\pi + 1 - \gamma_{\rm E}$):
\eqb
L(\mu) = {1\over 16\pi^2} \mu^{-2\epsilon}{1\over \hat \epsilon}.
\eqe
With the convention that an arbitrary coupling $k$ is written
as $k = k^{\rm r}(\mu) + \bar k\, L(\mu)$, the
following choices
render the Lagrangian Eq.~(\ref{eqn:ChHQL}) finite.
%
%
\begin{eqnarray}
\bar k_{10} &=& 6g \gamma {2\mu_0 \over f^2}
\\
\bar k_{11} &=& 6(g \gamma_1+g_1\gamma) {2\mu_0 \over f^2}
\\
\bar k_3 &=& (g \gamma_2+g_2\gamma){2\mu_0 \over f^2}
\\
\bar \epsilon_1 &=& 2\, g\, g_1 {m_0^2/3 \over f^2} 
\\
\bar \epsilon_2 &=& - 2\, g\, g_2 {m_0^2/3 \over f^2}
\end{eqnarray}
In addition, it is necessary to add the counter-term
\eqb
3g^2{m_0^2/3 \over f^2}
  \trd\left[\mybar H_a i v\cdot D_{ba} H_b\right]\,.
\eqe
The current Eq.~(\ref{eqn:currentfull}) is renormalized
with the choices
\begin{eqnarray}
\bar \rho_1 &=& \left[\frac12(1+3g^2)\rho_1
  + g(3g_1-g_2)\right]{m_0^2/3 \over f^2}, 
   \nonumber \\
\bar \rho_2 &=& \left[\frac12(1+3g^2)\rho_2
  + g g_2\right]{m_0^2/3 \over f^2}, 
   \nonumber \\
\bar \kappa_{10} &=& 6 g \gamma {\mu_0 \over f^2},
   \nonumber \\
\bar \kappa_{11} &=& \left[ g(3\gamma_1-\gamma_2)
	+ \gamma(3g_1-g_2)\right]{2\mu_0 \over f^2},
   \nonumber \\
\bar \kappa_3 &=&
   (g\gamma_2 + \gamma g_2){2\mu_0 \over f^2},
\end{eqnarray}
and it is also necessary to rescale $\alpha$:
\eqb
\alpha^{\rm r}(\mu) =
  \alpha\left[ 1 + \frac12(1+3g^2){m_0^2/3 \over f^2} L(\mu)\right].
\eqe
Finally, the mass terms in the Lagrangian must also be shifted
\begin{eqnarray}
\bar \lambda_3 &=& 16 g\gamma\lambda_2\, {2\mu_0 \over f^2}, 
	\nonumber\\
\bar \lambda_2 &=& 8g^2\, {m_0^2/3\over f^2}. 
\end{eqnarray}

\def\jvp#1#2#3#4{#1~{\bf #2}, #3 (#4)}
\def\PR#1#2#3{\jvp{Phys.~Rev.}{#1}{#2}{#3}}
\def\PRD#1#2#3{\jvp{Phys.~Rev.~D}{#1}{#2}{#3}}
\def\PRL#1#2#3{\jvp{Phys.~Rev.~Lett.}{#1}{#2}{#3}}
\def\PLB#1#2#3{\jvp{Phys.~Lett.~B}{#1}{#2}{#3}}
\def\NPB#1#2#3{\jvp{Nucl.~Phys.~B}{#1}{#2}{#3}}
\def\SJNP#1#2#3{\jvp{Sov.~J.~Nucl.~Phys.}{#1}{#2}{#3}}
\def\AP#1#2#3{\jvp{Ann.~Phys.}{#1}{#2}{#3}}
\def\PL#1#2#3{\jvp{Phys.~Lett.}{#1}{#2}{#3}}
\def\NuovoC#1#2#3{\jvp{Nuovo.~Cim.}{#1}{#2}{#3}}
\def\NPBPS#1#2#3{\jvp{Nucl.~Phys.~B~(Proc.~Suppl.)}{#1}{#2}{#3}}


\bibliographystyle{prsty}
\end{document}

\end{document}